\documentclass[prb,aps,groupedaddress,amsmath,twocolumn]{revtex4-1}
\usepackage[dvips]{graphicx}
\usepackage{dcolumn}
\usepackage{bm}
\usepackage{color}
\begin{document}

\title{Nuclear Quantum Effects Induce Metallization of Dense Solid Molecular Hydrogen} 

\author{Sam Azadi}
\email{s.azadi@imperial.ac.uk}
\affiliation{Department of Materials Science and Thomas Young Center,
 Imperial College London, London  SW7 2AZ, United Kingdom}
\author{ Ranber Singh}
\affiliation{Department of Physics, Sri Guru Gobind Singh College, Sector 26, 160019 Chandigarh, India}
\author{T.\ D.\ K\"{u}hne}
\affiliation{Department of Chemistry and Paderborn Center for Parallel
  Computing, University of Paderborn, Warburger Str.\ 100, D-33098 Paderborn,
  Germany Institute for Lightweight Design with Hybrid Systems, Warburger
  Str.\ 100, D-33098 Paderborn, Germany}

\date{\today}

\begin{abstract}
 We present an accurate computational study of the electronic structure 
 and lattice dynamics of solid molecular hydrogen at high pressure. 
 The band-gap energies of the $C2/c$, $Pc$, and $P6_3/m$ structures
 at pressures of 250, 300, and 350~GPa are calculated using the 
 diffusion quantum Monte Carlo (DMC) method. The atomic configurations
 are obtained from ab-initio path-integral molecular dynamics (PIMD)
 simulations at 300~K and 300~GPa to investigate the impact of zero-point
 energy and temperature-induced motion of the protons including 
 anharmonic effects. We find that finite temperature and nuclear quantum
 effects reduce the band-gaps substantially, leading to metallization of
 the $C2/c$ and $Pc$ phases via band overlap; the effect on the band-gap
 of the $P6_3/m$ structure is less pronounced. Our combined DMC-PIMD 
 simulations predict that there are no excitonic or quasiparticle energy
 gaps for the $C2/c$ and $Pc$ phases at 300~GPa and 300~K. Our results 
 also indicate a strong correlation between the band-gap energy and vibron
 modes. This strong coupling induces a band-gap reduction of more than
 2.46~eV in high-pressure solid molecular hydrogen. Comparing our DMC-PIMD
 with experimental results available, we conclude that none of the 
 structures proposed is a good candidate for phases III and IV of solid hydrogen.
\end{abstract}

\maketitle

\section {Introduction}
Determining the metallization pressure of solid hydrogen is one of the
great challenges of high-pressure physics.  Since 1935, when it was
predicted that molecular solid hydrogen would become a metallic atomic
crystal at 25~GPa \cite{W1935}, compressed hydrogen has been of huge
scientific interest.  Additional interest arises from the possible existence of
room-temperature superconductivity \cite{Ashcroft}, a metallic liquid
ground state \cite{Bonev}, and the relevance of solid hydrogen to
astrophysics \cite{Hemley,Ginzburg}.

Early spectroscopic measurements at low temperature suggested the
existence of three phases \cite{Silvera,Hemley}.  Phase I, which is stable up to
110~GPa, is a molecular solid composed of quantum rotors arranged in a
hexagonal close-packed structure. Changes in the low-frequency regions
of the Raman and infrared (IR) spectra imply the existence of phase II,
also known as the broken-symmetry phase, above 110~GPa.  The
appearance of phase III at 150~GPa is accompanied by a large
discontinuity in the Raman spectrum and a strong rise in the spectral
weight of the molecular vibrons.  Phase IV, characterized by the two vibrons
in its Raman spectrum, was discovered at 300~K and pressures above
230~GPa \cite{Eremets, Howie, Howie2}. Recently, another new phase has been
claimed to exist at pressures above 200~GPa and higher temperatures
(e.g. 480~K at 255~GPa) \cite{Howie3}. This phase is thought
to meet phases I and IV at a triple point, near which hydrogen retains
its molecular character. The most recent experimental results 
indicate that H$_2$ and hydrogen deuteride (HD) at 300~K and pressures
greater than 325~GPa transform into a new phase V that is characterized by
substantial weakening of the vibrational Raman activity\cite{Simpson}. Other features
include a change in the pressure dependence of the fundamental
vibrational frequency and partial loss of the low-frequency excitations.

Although it is very difficult to reach the static pressure of more than
400~GPa at which hydrogen is normally expected to metallize, some
experimental results have been interpreted as indicating metallization
at room temperature below 300~GPa \cite{Eremets}. However, other
experiments show no evidence of the optical conductivity expected of a
metal at any temperature up to the highest pressures explored
\cite{zha}.  Experimentally, it remains unclear whether or not the
molecular phases III and IV are metallic, although it has been suggested
that phase V may be non-molecular (atomic) and metallic \cite{Simpson}.
Metallization is believed to occur either via the dissociation of
hydrogen molecules and a structural transformation to an atomic metallic
phase \cite{samprl,Eremets}, or by band-gap closure within the
molecular phase \cite{MStadele,KAJohnson}. The mechanism and origin 
of band-gap closure are not well-known.  

The electronic structure of solid molecular hydrogen have mainly been
investigated by means of density-functional theory (DFT)-based methods
\cite{Pickard,Pickard2,Goncharov,Magdau,Naumov,Morales2013,Morales2014,JETP,singh,dft-fail}, 
as well as using the quasiparticle (QP) approach \cite{NJP,KAJohnson}. Although 
DFT-based methods are usually able to describe the crystal structures and
relative total energies of the relevant phases, their insufficiencies are
more apparent in the case of band-gap calculations \cite{JPPerdew}. To
obtain precise energy gaps, it is vital to go beyond mean-field theories 
and solve the many-electron Schr\"{o}dinger equation directly.  In this
work, we combine two sophisticated methods, diffusion quantum Monte Carlo
(DMC) and path-integral molecular dynamics (PIMD), to calculate the
excitonic and quasiparticle band-gaps of dense hydrogen at both zero 
and room temperature.

The DMC method is one of the most accurate known techniques for evaluating the total energies
of systems of more than a few tens of interacting quantum particles
\cite{DMC,Matthew1,AzadiO3}. It has been indicated recently that DMC can provide an
accurate description of the phase diagram of solid molecular hydrogen
\cite{Neil15}.  Although the DMC method was originally designed to study
the electronic ground state, it is also capable of providing accurate information
about excited states in atoms, molecules, and crystals
\cite{mitas,will,towler,Neil}. In general, DMC calculations of excitations in
crystals remain challenging because of the $1/N$ effect: The fractional
change in the total energy due to the presence of a one- or two-particle
excitation is inversely proportional to the number of electrons in the
simulation cell. Since large simulation cells are required to provide an
accurate description of the infinite solid, high-precision calculations
are necessary \cite{FiniteSizeDMC}.

Structures of crystalline materials are normally determined by x-ray or
neutron diffraction methods. These techniques are very challenging for
elements with low atomic numbers such as hydrogen, which is part of the reason the structures of
phases III and IV have remained uncertain. Fortunately, even though the crystal
structures are still unknown, optical phonon modes disappear, appear, or
experience discrete shifts when the crystal structure changes. It is,
therefore, possible to identify the transitions between phases using
optical methods.

The main input to any DMC calculation is the structure of the system
under study, which, in this case, is unknown. Hence, there is no option 
but to use structures predicted by mean-field theories such as DFT. It is now
generally accepted that DFT results for high-pressure hydrogen critically depends on
the choice of exchange-correlation (XC) functional
\cite{dft-fail,Morales2013,Morales2014}. This frustrating limitation may
be the main cause of the contradictions between different theoretical results \cite{JMcMinis,JMMcMahon}.  

In the present work, we carry
out a comprehensive study of the band-gap energy of high-pressure solid
molecular hydrogen at zero and finite temperature.  We concentrate on the smallest
band-gap, which may be direct or indirect, of the $C2/c$, $Pc$,
and $P6_3/m$ structures. These include all of the candidates suggested
by DFT calculations for phases III and IV
\cite{Pickard,Pickard2,Goncharov,Magdau,Neil15}.

\section {Computational Details}\label{CD}
We considered the $C2/c$, $Pc$, and $P6_3/m$
crystal structures of solid molecular hydrogen at P = 250, 300, and 350 GPa. 
According to previously conducted DFT simulations, the $C2/c$ and $Pc$ structures are the most favorable 
candidates for phase III and phase IV\cite{Pickard,Pickard2,Neil15}, respectively. 
All structures, which were afterwards used for our DMC simulations,
 were first fully relaxed at constant pressure at the DFT level. 
The latter calculations were carried out within the pseudopotential and
plane-wave approach using the Quantum Espresso suite of programs\cite{QS}.
All DFT calculations employed a dense 16$\times$16$\times$16 {\bf k}-point mesh,
 norm-conserving pseudopotentials, a plane-wave basis set 
with a cutoff of 100 Ry, as well as the Becke-Lee-Yang-Parr (BLYP) 
generalized gradient approximation to the exact XC functional\cite{BLYP}. 
In our recent work\cite{PRB16}, we have demonstrated that for studying the high-pressure 
solid hydrogen phase diagram, the BLYP XC functional
is more accurate than most other semi-local XC functionals. 
Independent studies of others have confirmed the superior accuracy of the employed 
BLYP XC functional for the description 
of thermodynamic properties of high-pressure hydrogen \cite{Clay16,Neil15,Morales2014}. 
Geometry and cell optimizations were performed using the Broyden-Fletcher-Goldfarb-Shanno quasi-Newton 
algorithm with a convergence
thresholds on the total energy and forces of 0.01 mRy and 0.1 mRy/Bohr,
respectively, to guarantee convergence of the total energy to better
than 1~meV/proton and the pressure to better than 0.1~GPa/proton.

Our DMC simulations were conducted using the \textsc{casino} QMC code \cite{casino} and
a trial function of the Slater--Jastrow (SJ) form,
\begin{equation}
\Psi_{\rm T}({\bf R})=\exp[J({\bf R})]\det[\psi_{n}({\bf r}_i^{\uparrow})]\det[\psi_{n}({\bf r}_j^{\downarrow})],
\label{eq6}
\end{equation}
where ${\bf R}$ is a $3N$-dimensional vector that defines the positions
of all $N$ electrons, ${\bf r}_i^{\uparrow}$ is the position of the
$i$th spin-up electron, ${\bf r}_j^{\downarrow}$ is the position of the
$j$th spin-down electron, $\exp[J({\bf R})]$ is the Jastrow correlation factor, whereas 
$\det[\psi_{n}({\bf r}_i^{\uparrow})]$ and $\det[\psi_{n}({\bf
  r}_j^{\downarrow})]$ are Slater determinants of spin-up and spin-down
one-electron orbitals, respectively. 
The one-electron orbitals were also obtained from the DFT 
calculations using the plane-wave Quantum Espresso code
\cite{QS}. However, to guarantee converge to the complete basis
set limit, a rather large basis set cutoff of 200~Ry was chosen\cite{sam}. 
Thereafter, the plane-wave orbitals were transformed into a
blip polynomial basis\cite{blip}.

As is customary, our QMC calculations were carried out using finite simulation cells 
subject to periodic boundary conditions which introduce finite-size (FS)
errors. An important contribution to the FS errors in our calculations 
arises from the treatment of the Coulomb potential energy. The Coulomb 
interaction is inconsistent with the periodicity of the simulation cell
and has to be replaced by the Ewald interaction, which is the Green's 
function of Poisson's equation subject to periodic boundary conditions. 
Unlike the Coulomb interaction, the Ewald interaction depends on the size 
and shape of the simulation cell. Therefore, we employed canonical twist averaging
to correct single-particle FS errors which are analogous to 
{\bf k}-point sampling errors in DFT\cite{tav}. 
Specifically, we performed twist-averaged DMC calculations based on eight randomly 
 chosen twists at three different system sizes for each phase and volume.
 We used 192, 432, and 648 atoms in our simulation cells for the $C2/c$ phase,
 which has 24 atoms in the primitive unit cell, 192, 384, and 576 atoms for
 the $Pc$ structure with 48 atoms in the primitive unit cell, and 28, 288,
 and 768 hydrogen atoms for the $P6_3/m$ phase that has 16 atoms in the primitive unit cell.

We linearly extrapolated the twist-averaged energy per atom to 
the thermodynamic limit in order to correct for many-body FS errors, which are due to the long range
of both the Coulomb interaction and the two-body correlations. We verified that the linear fitting is accurate, 
because the finite-size error in the energy per atom deceases 
as $1/N$, where $N$ is the number of particles in the simulation cell. 
 
Our Jastrow correlation factor consists of a polynomial one-body electron-nucleus (1b),
two-body electron-electron (2b), three body electron-electron-nucleus (3b),
as well as plane-wave expansions in the electron-electron separation ($p$)\cite{pterm}.
We found that the latter $p$ term makes significant improvements to the wave function  
and the variational energy, since it describes the long-range correlation term in the Jastrow, which plays
an important role in our calculations.
The parameters within the Jastrow were optimized by means of variance
minimization at the variational quantum Monte Carlo (VMC) level\cite{varmin1,varmin2}.

The QP energy gap is defined as
\begin{equation}
\label{eq_qp}
\Delta_{\rm qp}= E_{N+1}+E_{N-1}-2E_0, 
\end{equation}
where $E_{N+1}$ [$E_{N-1}$] is the many-body
total energy of the system after an electron has been added [removed form] the system, 
while $E_0$ is the ground-state energy.
Our calculations of $\Delta_{\rm qp}$ were performed at the $\Gamma$-point 
of the supercell Brillouin zone, equivalent to a mesh of $k$-points in the primitive
Brillouin zone. Technical details of the QP calculations are given in
our recent work\cite{PRB16}.
We created excitonic states by promoting an electron from a
valence-band orbital into a conduction-band state with the same
wave vector. 
However, the so created states do not cover all excitonic effects since no 
configuration interactions are taken into account.
The absorption gap to an excitonic state at $\Gamma$ is 
\begin{equation}
\label{eq1}
 \Delta_{\rm exc} = E^{\prime} - E_{0},
\end{equation}
where $E^{\prime}$ and $E_{0}$ are the total energies of the 
first excited and the ground state as obtained by DMC.  We
found that the difference between the triplet and singlet excitonic gaps
is small in the systems studied here. For example, in the case of the
$C2/c$ structure at 250~GPa, the singlet-triplet splitting is 0.2(1)
eV. We focus, therefore, on singlet excitonic absorption gaps.

Finite-temperature second generation Car-Parrinello PIMD simulations were 
performed to account for harmonic and anharmonic zero-point motion (ZPM) \cite{PRL, Wires, PRE}. 
The combined path-integral generalized Langevin (PIGLET) scheme of Ceriotti et al. 
was employed here \cite{Ceriotti12}, as implemented in the i-Pi wrapper\cite{iPi},
 to reduce the number of computationally expensive imaginary-time replicas.

Eventually, convergence was achieved with as few as six imaginary-time replicas. 
All of the simulations were performed using the isobaric-isothermal (NPT)
ensemble at 300 K and 300 GPa. The interatomic forces were computed at the DFT level using the 
Quantum Espresso suite of programs and a plane-wave 
cutoff of 680 eV. The simulations consisted of 96 protons for the $C2/c$ and $Pc$ phases 
and 128 atoms for the $P6_3/m$ structure. 
The first Brillouin zone was sampled using a $2\times2\times2$ Monkhorst-Pack k-point mesh
\cite{prb76}. In each case, 25 statistically independent configurations,
which are separated by 1~ps, were considered to compute the 
associated ensemble averages at the DMC level. Since for each PIMD configuration
 the DMC energy is obtained based on eight randomly chosen twists\cite{NJP},
 in total $3\times25\times8=600$ DMC simulations have been performed to obtain reliable results. 

\section {Results and discussion}\label{RD}
\begin{figure}
\includegraphics[scale=0.5]{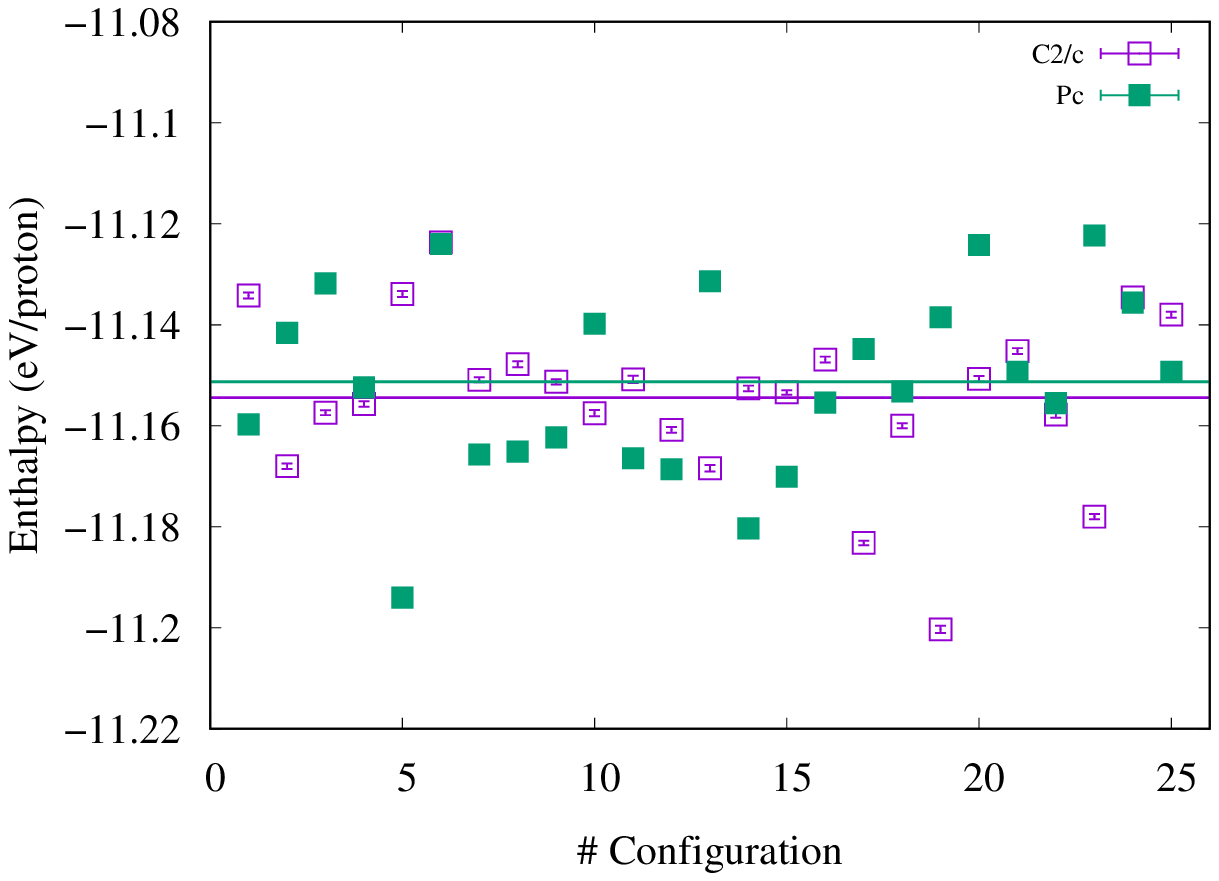}\\
\includegraphics[scale=0.5]{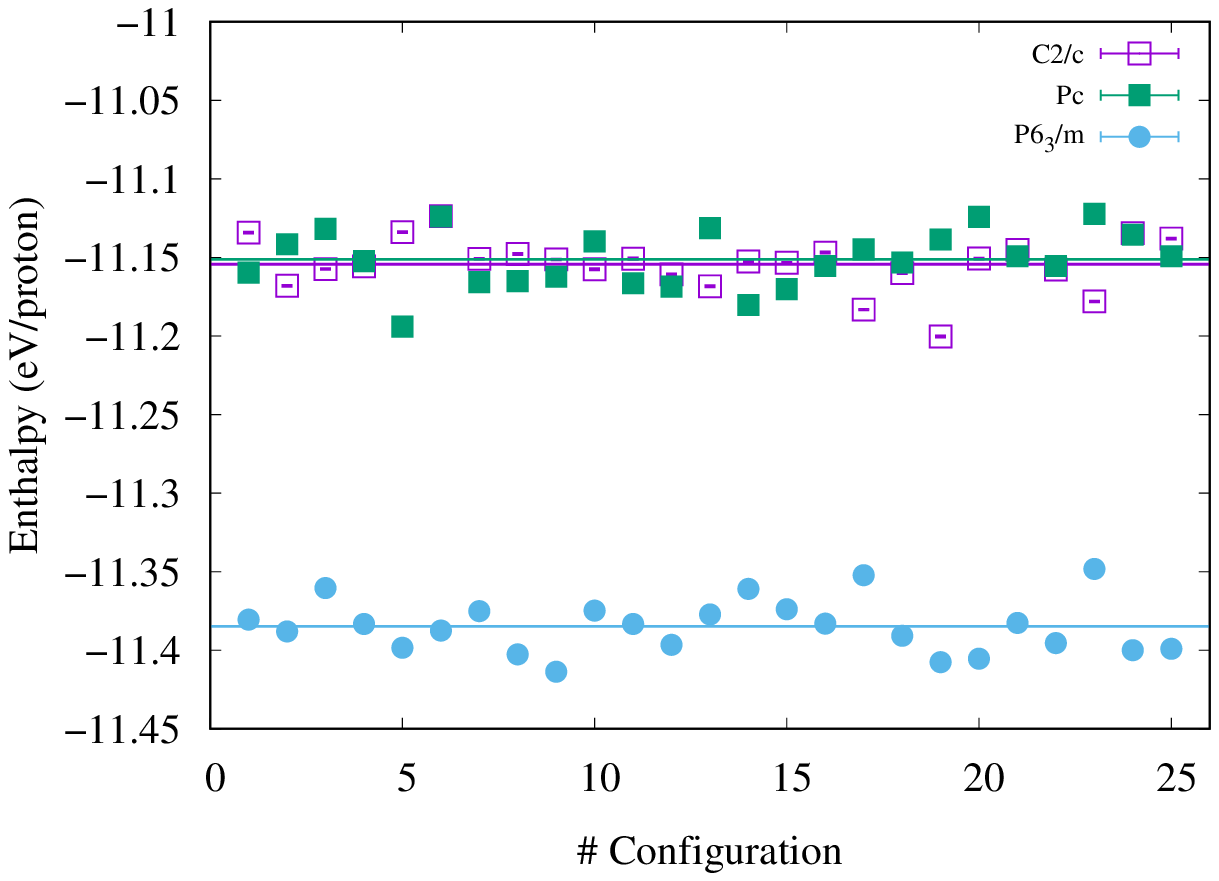}\\
\caption{\label{EDMC_PIMD} DMC enthalpies of 25 statistically independent configurations 
obtained by \textit{ab-initio} PIMD simulation at P= 300~GPa and T=300~K. 
The ensemble averaged energy is illustrated by a flat line.}
\end{figure}

The free energy difference between two phases at constant temperature and pressure is
$\Delta G = \Delta H - T \Delta S$, where $\Delta H = \Delta E + P \Delta V$. 
Herein, $G, H, T, S, E, P$ and $V$ are the Gibbs free energy, enthalpy, temperature, entropy, 
internal energy, pressure and volume of the system. 
We calculated the ensemble average of the enthalpy 
$\langle H \rangle = {N}^{-1} \Sigma_{i=1}^{N} H_{i}$, where $H_i$ is the enthalpy of the system at each 
configuration. Thus, the enthalpy difference between two phases can be defined as 
$\langle \Delta H \rangle = \langle H_2 \rangle - \langle H_1 \rangle$.  
In Fig.~\ref{EDMC_PIMD}, $\langle H \rangle$ for the $C2/c$, $Pc$, 
and $P6_3/m$ phases at room temperature and 300~GPa are shown. 
The resulting difference $\langle \Delta H \rangle$ between the $C2/c$ and $Pc$ structures is about 
2(1)~meV/atom, while it is 0.23(2)~eV/atom between the $P6_3/m$ and the two other phases. 
However, at constant temperature, $\Delta S$ between the $C2/c$ and $Pc$ phases is   
less than 10~meV/atom \cite{Neil15}. 
This suggests that $\Delta S$ is the 
dominant term within $\Delta G$ for the $C2/c$ to $Pc$ transition and that 
this is an entropy driven transition. 
Assuming that $\Delta S$ is relatively similar for all considered structures, 
this also suggests that, at room temperature and 300~GPa, the Gibbs free energy of $P6_3/m$ is much 
lower than that of the $C2/c$ and $Pc$ phases, respectively.

\subsection{Electronic Density of States}
\begin{figure}
\includegraphics[scale=0.5]{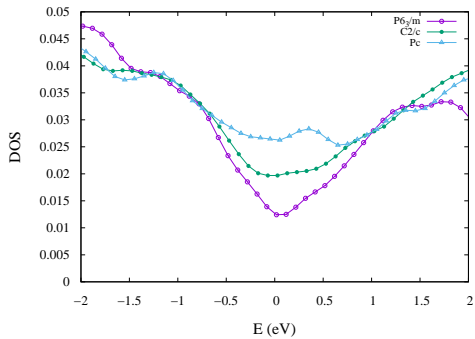}\\
\caption{\label{AVEDOS} Ensemble averaged DOS per particle as obtained by \textit{ab-initio} PIMD 
simulations at P= 300~GPa and T=300~K for the $C2/c$, $Pc$, and $P6_3/m$ 
phases. The Fermi energy is located at E=0. All energies and DOS are in eV and eV$^{-1}$, respectively.}
\end{figure}

We investigated the electronic structure of the $C2/c$, $Pc$, and $P6_3/m$ by 
computing the averaged DOS at room temperature and 300~GPa, which is shown in Fig.~\ref{AVEDOS}. 
As can be seen, the electronic DOS at Fermi level of the $P6_3/m$ phase is smaller than 
that of the $C2/c$ and $Pc$ structures. Nevertheless, at the DFT level, all of the considered phases obey 
metallic behavior.

\subsection{Band-gap Energy}
In a recent paper we reported our QP and 
excitonic band-gap results at the nuclear ground state, where the
ZPM and temperature-induced vibrations of the 
protons are neglected \cite{PRB16}. Therein, we found that the exciton 
binding energy is smaller than 100~meV/atom and
that our DMC QP and excitonic band-gaps are within
error bars to one another. Therefore, in this work,
we do not attempt to distinguish the excitonic from the QP band-gap.
Instead, by calculating the static-nucleus energy band-gaps using DMC, 
we quantify the many-body corrections to the band
structure that are absent in the mean-field DFT approach. Although these 
contributions are relatively independent of lattice vibrations and
temperature, it is well known that nuclear quantum effects are significant in
hydrogen-rich systems and greatly affect the metallization of 
high-pressure solid hydrogen \cite{Morales2013}.  Assuming the validity of the
Born-Oppenheimer approximation, the full electron-nuclear wave function
$\Psi({\bf R}, {\bf d})$ can be approximated in the form $\Phi({\bf R} |
{\bf d}) \chi({\bf d})$, where $\Phi({\bf R} | {\bf d})$ is a function
of the positions ${\bf R} = ({\bf r}_1,{\bf r}_2,\ldots,{\bf r}_N)$ of
the $N$ electrons in the supercell at fixed nuclear positions ${\bf d}$,
while $\chi({\bf d})$ is the nuclear wave function. The measured
band structure is then an average of the band structures calculated from
$\Phi({\bf R} | {\bf d})$, which are weighted according to the nuclear probability
density. We sampled the nuclear probability density by performing 
finite-temperature PIMD simulations using DFT within the generalized
gradient approximation in order to take harmonic and anharmonic nuclear quantum and
finite-temperature effects into account.

It is not straightforward to measure the band-gap at pressures
greater than 300~GPa, but experimental results
suggest that solid hydrogen is indeed an insulator
in this pressure range\cite{Goncharov,zha,Howie}.
It is, therefore, reasonable to assume that the
$C2/c$, $Pc$, and $P6_3/m$ structures all have
non-zero band-gaps at 300~GPa and 300~K. Based on this assumption, we treat each PIMD
configuration as an insulator and calculate the DFT band structure of
that configuration by occupying the same number of one-electron 
DFT orbitals at every wave vector in the simulation cell Brillouin zone. 
Using this protocol, we found that the highest occupied
and lowest unoccupied states belong to different {\bf k}-points
and that the difference between them, which we denote as the
dynamic gap, is negative for almost every nuclear configuration
sampled. This implies that, at the DFT level, the nuclear quantum and
finite-temperature effects render all the studied structures
metallic.

\begin{figure}
\includegraphics[width=0.4\textwidth]{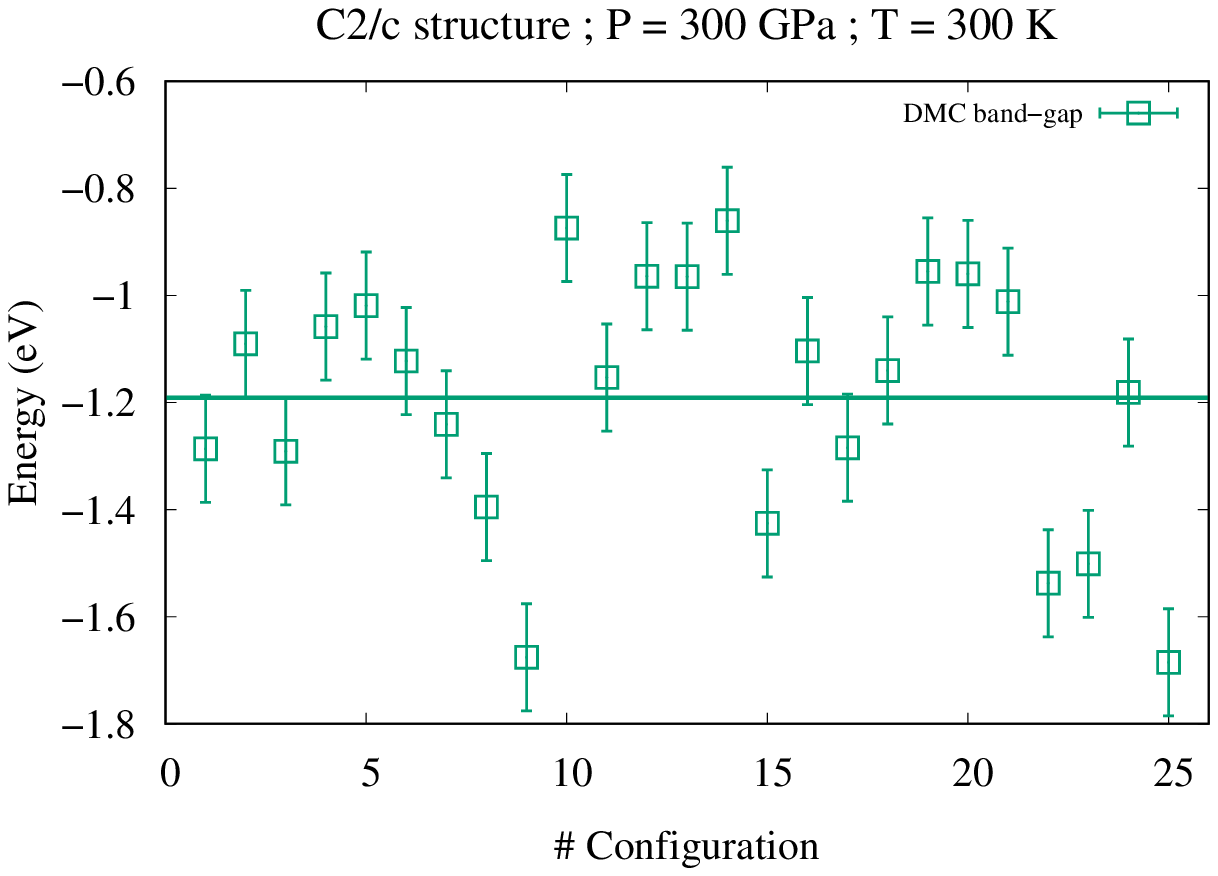} \\
\includegraphics[width=0.4\textwidth]{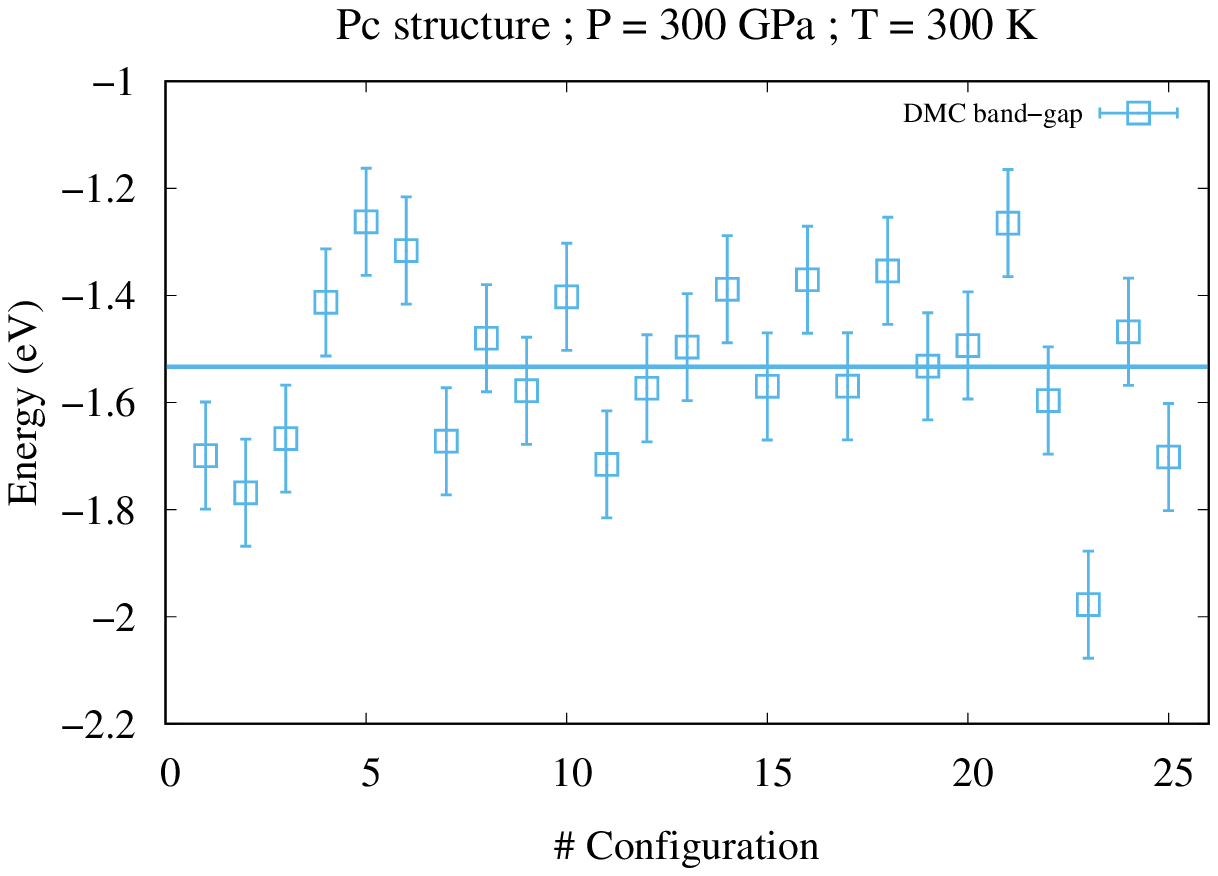}\\
\includegraphics[width=0.4\textwidth]{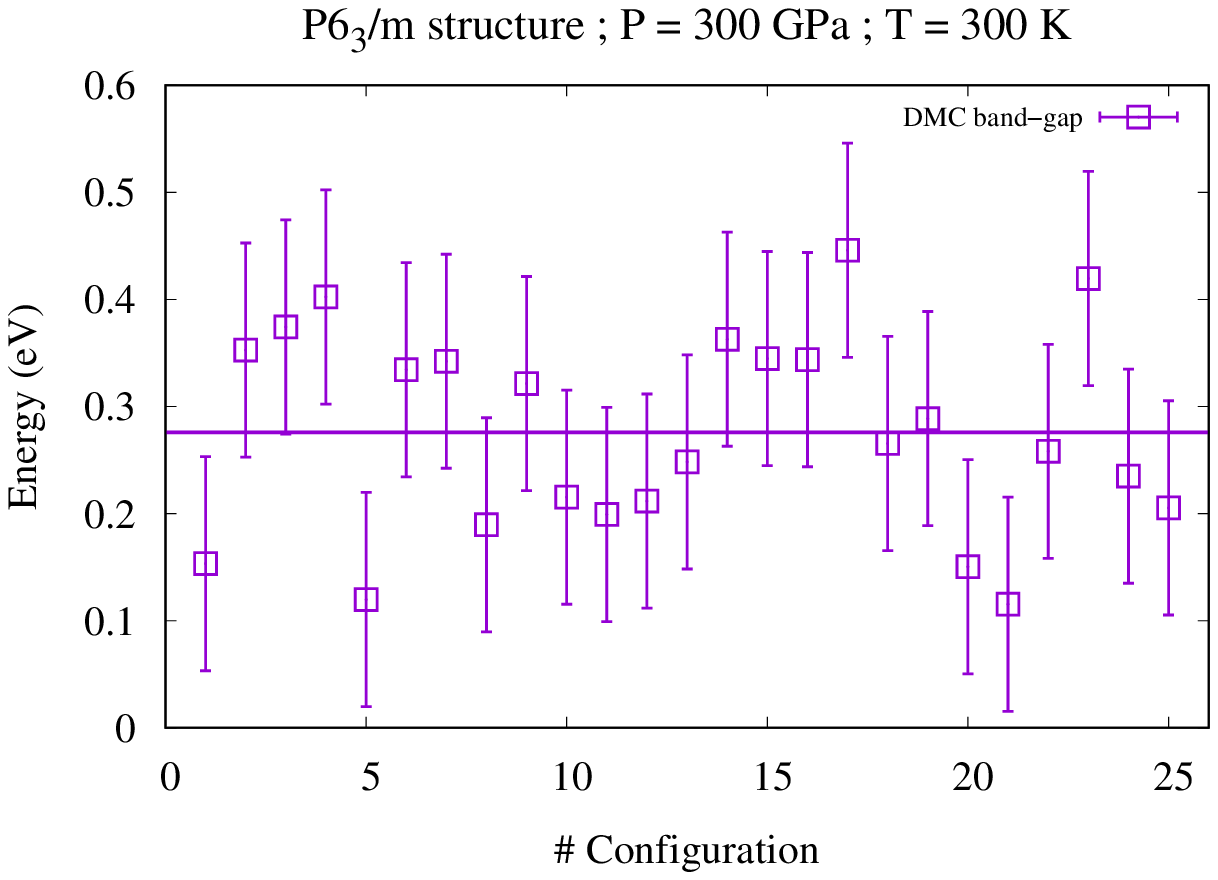}
\caption{\label{PIMD_gap} Energy band-gaps of the $C2/c$, $Pc$, and $P6_3/m$ 
structures at room temperature and a pressure of 300~GPa. The horizontal lines 
denotes the corresponding ensemble averaged band-gap energies. 
A DMC scissor operator is used to correct the DFT energy band-gaps.
The statistical uncertainties within the band-gap as obtained by 
the present DMC calculations are indicated be vertical bars. The occurence of negative 
band-gaps is due to the fact that the valence band maximum (VBM) and 
conduction band minimum (CBM), which occur at different $\bf k$-points, 
do overlap.}
\end{figure}

To correct for the underestimation of the dynamic gap by DFT, we introduce
a {\it scissor operator}.  The dynamic DFT band-gap is increased from
$E_{g}^{DFT}$ to $E_{g}^{DFT} + \delta_{sci}$, where $\delta_{sci} = E_{g}^{DMC} -
E_{g}^{DFT}$ is calculated for a perfect crystalline supercell at the
static level\cite{PRB16}. 
The scissor operator depends on 
the crystal structure and the pressure applied. However, a detailed study and 
the specific values of $\delta_{sci}$ for all the studied 
structures at P = 250, 300, and 350~GPa are reported in our recent 
work\cite{PRB16}. The resulting scissor-corrected dynamic 
band-gaps of the $C2/c$, $Pc$, and $P6_3/m$ structures at room
temperature and a pressure of 300~GPa are shown in Fig.~\ref{PIMD_gap}. 
We find that the harmonic and
anharmonic ZPM and finite-temperature contributions are substantial, and
that the $C2/c$ and $Pc$ structures remain metallic even after the
scissor correction has been applied. The scissor-corrected static 
band-gaps of the $C2/c$, $Pc$, and $P6_3/m$ structures at 300~GPa,
which are calculated by DMC, are 2.3(2), 2.4(2), and 2.8(2) eV,
respectively \cite{PRB16}. The inclusion of nuclear quantum
and finite-temperature effects lowers the $C2/c$, $Pc$, and $P6_3/m$
band-gaps by as much as 3.4(2), 3.8(2), and 2.5(2)~eV, respectively, relative to
static-nucleus results. 

\subsection*{Discussion}
Our results indicate a strong structure dependence of 
the protons ZPM and thermal contributions. The
$C2/c$ and $Pc$ crystal structures have weakly bonded graphene-like layers,
while three-quarters of the H$_2$ molecules in the $P6_3/m$ phase lie flat 
within the plane and a quarter lie perpendicular to
the plane. The centers of the molecules fall on a slightly distorted
hexagonal close-packed lattice \cite{Pickard}.  The quantum and thermal
vibrational motions increase the intermolecular interactions and weaken
the intramolecular bonding, an effect that appears to be more
significant for the $C2/c$ and $Pc$ structures than for the $P6_3/m$
phase. The high structural flexibility of phase IV at pressures of
250--350~GPa and temperatures of 300--500~K has also been observed in
a previous {\it ab initio} variable-cell MD simulations, which found
that the protons in the graphene-like layers can readily transfer to
neighboring molecular sites via a simultaneous rotation of
three-molecule rings \cite{HLiu}. Our results confirm that this behavior has a
strong effect on the band-gaps of the $C2/c$ and $Pc$ structures.

According to our band-gap energy results, the $C2/c$ and $Pc$
structures, thought to be the best candidates for phases III and IV
of solid molecular hydrogen \cite{Neil15}, are metallic at room temperature and 300~GPa, in
disagreement with most of the experimental evidence. Another possible
crystal structure, $Pbcn$, entails two different layers of graphene-like
three-molecule rings with elongated and unbound H$_2$
molecules \cite{Pickard}.  Assuming that the reduction of the band-gap 
due to the proton motion in $Pbcn$ is similar to that in $Pc$, we would also
expect $Pbcn$ to be metallic at room temperature and 300~GPa. There is
some experimental evidence of metallization of high-pressure hydrogen at
room temperature and pressures around 300~GPa \cite{Eremets}, but the
onset of metallic behavior was thought to be accompanied by a
first-order structural transition, presumably into a monatomic liquid
state. 

There are experimental band-gap results up to a pressure of 350~GPa\cite{Goncharov,zha,Howie}. 
The optical transmission spectrum of phase IV
shows an overall increase of absorption and a band-gap reduction 
to 1.8~eV at 315~GPa \cite{Howie}. According to these results,
metallization of solid hydrogen at and below room temperature should
happen at pressures above 350~GPa. This scenario appears to contradict our
band-gap energy results for the $C2/c$ and $Pc$ phases. Assuming that the experimental
measurements of the band-gaps are accurate, our analysis implies that the $C2/c$
and $Pc$ phases are unlikely candidates for high-pressure
solid molecular hydrogen at room temperature and 300~GPa. Instead, it is more 
likely that the $C2/c$ and $Pc$ structures become semi-metallic at
pressures between 250 to 300~GPa.

The dynamic band-gap of the $P6_3/m$ phases at room temperature and 300~GPa is 
larger than the that of the $C2/c$ and $Pc$ structures. The static DMC band-gap of the $P6_3/m$
phase is also larger than that of the other layered-structures. In addition, the band-gap 
reduction, due to ZPM and temperature contributions, is in the $P6_3/m$ phase $\sim$1~eV smaller 
than in the two other structures studied. 
Experimental Raman spectra, however,
suggest that phase IV consists of two different graphene-like sheets and
unbound H$_2$ molecules \cite{Howie}. The metallic nature of the $C2/c$
and $Pc$ structures leads us to doubt that any graphene-like structure
with weak interactions between layers remains an insulator at room
temperature and 300~GPa, so this interpretation of the Raman data is
difficult to reconcile with the existence of a band-gap. 
It seems likely that the true structures of phase III and
phase IV of high-pressure hydrogen remain unknown \cite{Howie3}.

\begin{figure}
\includegraphics[width=0.4\textwidth]{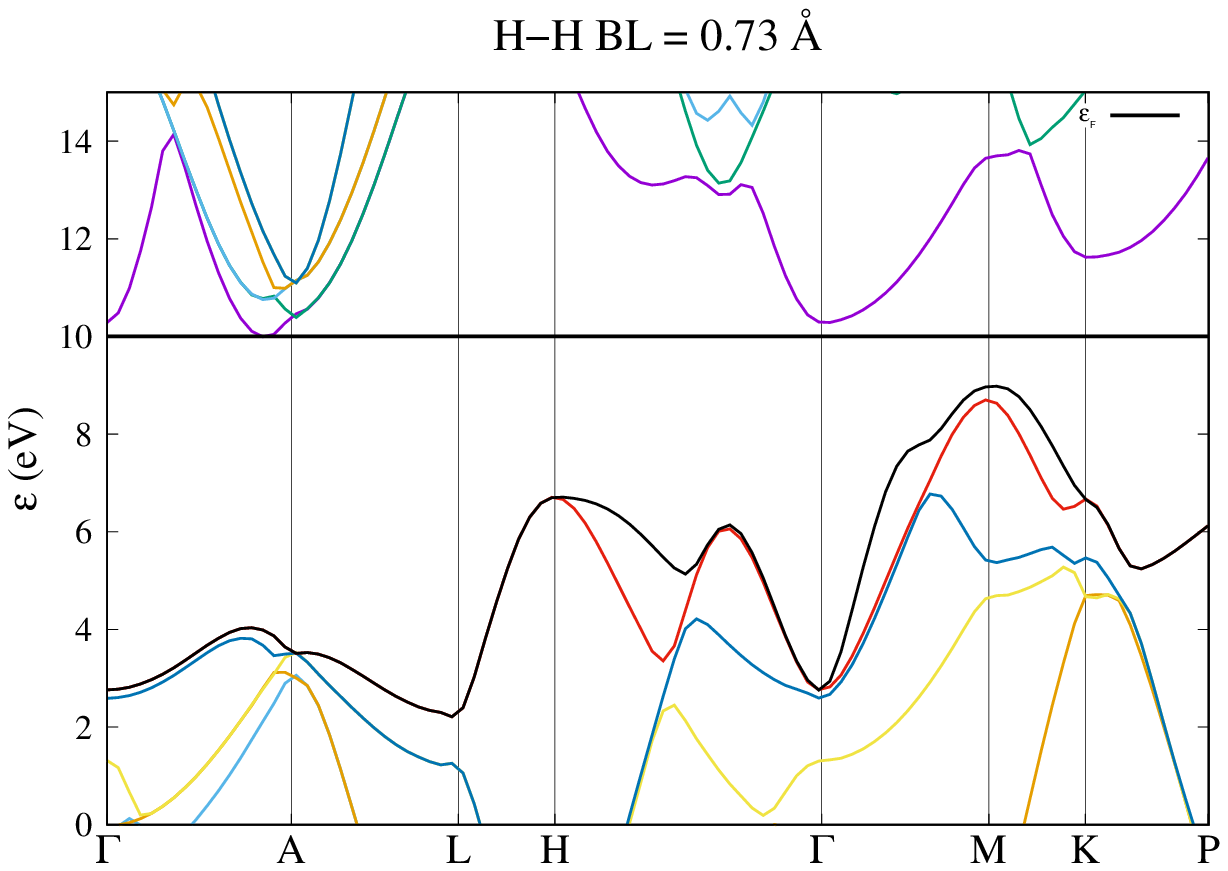}\\
\includegraphics[width=0.4\textwidth]{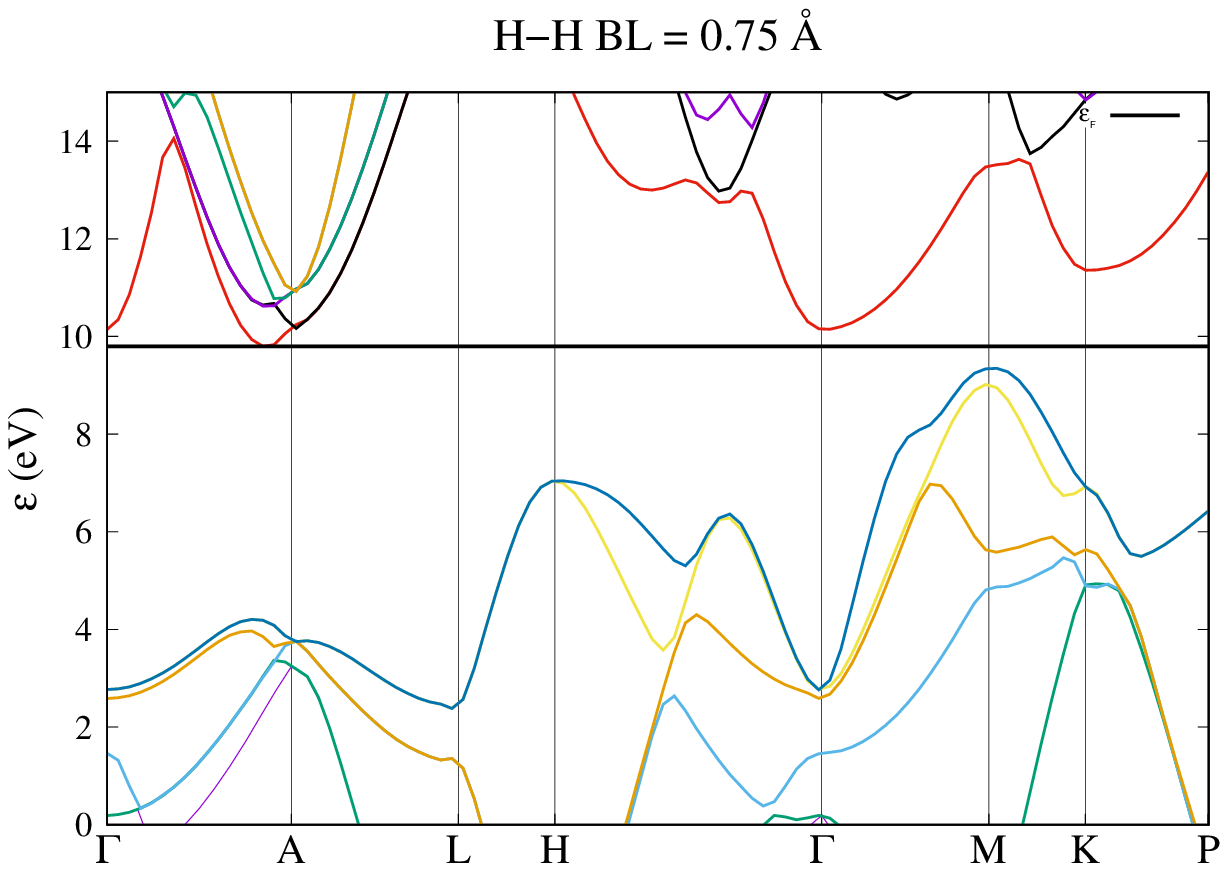}
\caption{\label{P6_HH} Electronic band structure of the $P6_3/m$ 
structure at fixed volume. The DFT results are obtained for two 
intramolecular H--H  bond-lengths (BL) of 0.73 and 0.75 \AA, respectively. 
 The location of the Fermi level is indicated by the black horizontal line.}
\end{figure}

To gain a deeper understanding of this problem, we studied the correlation between 
the electronic band structure and the intramolecular H--H bond-length. For that purpose, we performed 
band structure calculations for the $P6_3/m$ phase, which has the largest band-gap, 
at fixed volume and two intramolecular H--H bond-lengths (BL) of 0.73 and 0.75 \AA 
(Fig.~\ref{P6_HH}), respectively. 
We found a strong coupling between the highest-occupied state 
and the intramolecular H--H BL. The gradient of the energy gap 
with respect to the intramolecular BL is $\sim 27.3$ eV/\AA, and is independent 
of XC functional \cite{PRB16}. Since the root mean vibration amplitude 
of a $H_2$ molecule due to the vibron modes is $\sim 0.09$ \AA
\cite{Labet,Labet2,Labet3,Labet4}, the band-gap reduction caused 
by ZP vibrations is $\sim 2.46$ eV, which is in excellent agreement with the 
2.5(2)~eV band-gap reduction obtained by our PIMD calculations. 

\section{Conclusion}\label{con}
To summarize, based on combined \textit{ab-initio} PIMD and DMC simulations, 
we find that finite-temperature and nuclear quantum effects leads to a 
band-gap reduction of at least 2.5~eV. Since this renders all candidate 
structures we have considered here metallic, which is in contrast with 
available experimental measurements, we conclude that none of the proposed
 structures are likely candidates for phase III and IV of solid molecular hydrogen.

The authors would like to thank the Gauss Center for Supercomputing (GCS) for 
providing computing time through the John von Neumann Institute for Computing (NIC)
 on the GCS share of the supercomputer JUQUEEN at the J\"ulich Supercomputing 
Centre (JSC). Additional computing facilities were provided through the DECI-13 
PRACE project ÔÔQMCBENZ15ÕÕ and the Dutch national supercomputer Cartesius. 
This project has received funding from the European Research Council (ERC)
 under the European Union's Horizon 2020 research and innovation programme (grant agreement No 716142).

%\newpage
%\begin{thebibliography}{99}

\end{document}